\documentclass[12pt,aps,pre,twocolumn,showpacs,groupedaddress,longbibliography
,superscriptaddress
]{revtex4-1}

\usepackage{graphicx}
\usepackage{bm}

\begin{document}

\title{Towards room-temperature superfluidity of exciton polaritons in an optical microcavity with an embedded MoS$_2$ monolayer}

\author{German V. Kolmakov\footnote{Corresponding author: gkolmakov@citytech.cuny.edu}}
\affiliation{Physics Department, New York City College of Technology, The City University of New York, Brooklyn, New York 11201}

\author{Leonid M. Pomirchi}
\affiliation{Physics Department, New York City College of Technology, The City University of New York, Brooklyn, New York 11201}

\author{Roman Ya. Kezerashvili}
\affiliation{Physics Department, New York City College of Technology, The City University of New York, Brooklyn, New York 11201}
\affiliation{The Graduate School and University Center, The City University of New York, New York, New York 10016}

%\ociscodes{(190.4400)   Nonlinear optics, materials; (240.5420)   Polaritons.}

%\doi{\url{http://dx.doi.org/10.1364/ao.XX.XXXXXX}}

\begin{abstract}
By  considering driven diffusive dynamics of exciton polaritons in an optical microcavity with an embedded molybdenum disulfide monolayer,
we determine experimentally relevant range of parameters at which room-temperature superfuidity can be observed.
It is shown that the superfluid transitions occurs in a trapped polariton gas  at laser pumping power $P>600$ mW and and trapping potential strength
$k > 50$ eV/cm$^2$. We also propose a simple analytic model that provides a useful  estimate for the polariton gas  density, which 
enables one to
determine the conditions for observation of room temperature polariton superfluidity. 
\end{abstract}

%\setboolean{displaycopyright}{true}

\maketitle
%\thispagestyle{fancy}

%\ifthenelse{\boolean{shortarticle}}{\ifthenelse{\boolean{singlecolumn}}{\abscontentformatted}{\abscontent}}{}

\section{Introduction}
In the past decade, physics of  polaritons, a quantum superposition of excitons and cavity photons,  attracts significant 
attention due to fundamental interest to macroscopic collective quantum phenomena in this system \cite{Carusotto:13} as well as 
due to potential applications in photonics and optoelectronics \cite{Gibbs:11}. 
Thanks to their small effective mass, polaritons
demonstrate the  Bose-Einstein condensation (BEC) and superfluidity at temperatures about six orders higher than that for
ultra-cold atomic gases \cite{Deng:02,Kasprzak:06,Balili:07,Utsunomiya:08,Amo:09a}. 
 Numerous applications of  
polariton condensates  in optical computing \cite{Bajoni:08,Menon:10,Gao:12,Ballarini:13,Berman:14}, nonlinear 
interferometry \cite{Sturm:14}, novel light sources \cite{Deng:02a,Malpuech:02a,Szymanska:02,Bajoni:07,Christopoulos:07}, 
and  atomtronics \cite{Restrepo:14}
have recently been proposed.

Transition-metal dichalcogenide monolayer crystals  provide unprecedented opportunity to transfer
 these achievements to the room temperature scale.  
In contrast to another near-perfect two-dimensional material,  graphene,
single-layer molybdenum and tundsen-based transition metal dichalcogenides 
are direct-zone semiconductors with the sizable bandgap
$E_g\sim 1-2$ eV \cite{Das:14}.
Extremely  large light-matter interaction strength (the Rabi splitting $\sim 50$ meV)  \cite{Liu:15} and  large
exciton binding energy $\sim 0.3-0.9$ eV  in monolayer 
transition metal dichalcogenides \cite{Qiu:13,Chernikov:14} make these materials exceptional candidates  
 for observation of room temperature quantum collective phenomena.
Observation of polaritons in a microcavity with embedded molybdenum disulfide (MoS$_2$) 
monolayer at the room temperature has recently been reported in Ref.~\cite{Liu:15}. 
 
However, it is  {\it a priory} unclear  if the conditions for a superfluid transition can be satisfied in actual experiments
for a polariton gas in cavities with transition-metal dichalcogenide monolayers at room temperatures. 
One of the important factors is significant temperature 
fluctuations in the system 
that results in spreading of a polariton cloud in a trap and thus, in a drop of the polariton density.
We note that temperature spreading  is not essential for conventional low-temperature polariton experiments 
with semiconductor microcavities where the leading factors are quantum fluctuations 
and polariton-polariton interactions ~\cite{Berman:08}. 

 The goal of this work is to determine a range of experimentally relevant parameters,
for which superfluidity of a polariton gas can be achieved at room temperatures
in an optical microcavity with an embedded MoS$_2$ monolayer.
Specifically, by considering the quasiclassical diffusive dynamics of a trapped
polariton gas 
we determined the values of the trapping potential strength 
and laser pump power, for which the room temperature  
polariton superfluidity can be observed.
 
 \begin{figure}[t]
 \includegraphics[width=8cm]{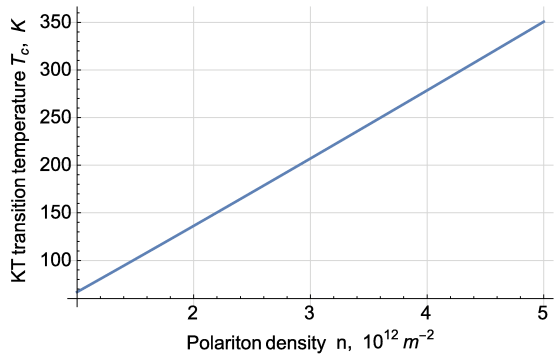}
 \caption{\label{fig:tc}
Temperature of the Kosterlitz-Thouless (KT) phase transition, $T_c$, as a function of the polariton gas density $n$ 
estimated from Eq. (\ref{eq:tc}) where the cutoff length $r_0$ is set equal to the exciton Bohr radius 
in molybdenum disulfide $a_B \approx 1$ nm and the polariton mass is $m = 3.14 \times 10^{-5} m_0$, see text.
It is seen that the KT transition at the room temperature $T=300$ K occurs at the density 
$n \approx 4.2 \times 10^{12}$ m$^{-2}$.
}
\end{figure}

\section{Superfluid phase transition in a  polariton gas}
A homogenous two-dimensional Bose gas can only transits to a Bose condensed state, in which the occupation 
of the lowest energetic level is macroscopically large, at zero temperature since the temperature fluctuations 
destroy the long-range order in the system.
In contrast, an ideal trapped quantum Bose gas undergoes the BEC transition at a finite temperature 
$T_0 = \hbar (\pi k_B)^{-1}\sqrt{6k N/m}$ where $k$ is the effective trapping potential strength, $N$ is the total number of particles, and 
$m$ is the particle mass \cite{Bagnato:91}. 
At low temperatures $T \ll T_0$ virtually all particles occupy the lowest energetic level that is, 
a condensate is formed \cite{Dalfovo:99}.
However, for a Bose gas of interacting particles, the condensate can be significantly depleted due to mutual particle scattering:
the condensate fraction for a system of $N \gg 1$ particles falls as  $N_c/N \propto \log(N)/N^{1/2}$ \cite{Holzmann:07}.
In effect, despite $T_0$ for an ideal gas of polaritons can be estimated as a few hundred K because of a small mass $m$, 
the actual BEC transition temperature of a trapped polariton gas lies in the helium temperature range~\cite{Berman:08}.

Nevertheless, a Bose system with interactions can transit to a superfluid state even if the condensate fraction is small or  absent et all.
The famous example is superfluid $^4$He (He-II), in which the occupation of the lowest energetic
level is only $\sim 10$\%, and $\sim 90$\% of particles are ``pushed'' out of the condensate to higher energetic levels 
due to strong interatomic interactions \cite{Atkins:59}. The relation between the Bose condensation and superfluidity is 
delicate and is discussed in a number of 
publications (e.g., \cite{Dalfovo:99,Yukalov:11,Berman:11a}). 
The superfluid transition in a two-dimensional system with interactions is a Kosterlitz-Thouless (KT)  phase transition, which is related to pairing of 
quantized vortices \cite{Kosterlitz:73}. The KT transition temperature for a  gas of interacting polaritons
is given by the following equation~\cite{Lozovik:06}
\begin{equation}
T_c = \frac{2 \pi \hbar^2 n}{m \log \log (1/n r_0^2)} \label{eq:tc}
\end{equation}
where $n$ is the gas density and $r_0$ is the small-scale cutoff length. The  dependence (\ref{eq:tc}) for polaritons in a cavity with embedded 
molybdenum disulfide is shown in Fig.\ \ref{fig:tc}. It is seen in Fig.\ \ref{fig:tc} that a polariton 
gas  can undergo a superfluid transition at the room temperature $T=300$ K if the gas density is 
high enough, $n>n_c \approx 4.2 \times 10^{12}$ m$^{-2}$. At $n\sim n_c$, the
thermal wavelength of polaritons  $\lambda_T = 2^{1/2}  \pi \hbar / (m k_B T)^{1/2} \approx 1.3$ $\mu$m
is about five times larger than the inter-particle distance $r_s = (\pi n)^{-1/2} \approx 0.28$ $\mu$m and thus, 
quantum collective effects are important at densities above $n_c$. However, the gas parameter is small, 
$n_c a_B^2 \sim 4 \times 10^{-6}$,
that justifies the application of the weakly interacting gas  model (\ref{eq:diffusion}), see the next sections.

In what follows we identify the range of experimental parameters at which the KT superfluid transition can be observed
in a trapped polariton gas in microcavities with an embedded transition-metal dichalcogenide monolayer.
In our consideration, we focus on microcavities with an embedded molybdenum disulfide
monolayer that is, on a system where room-temperature polaritons have  recently been 
 observed \cite{Liu:15}. 

\section{Simulation Method}
The dynamics of a normal  polariton gas at the scales much larger than the polariton mean free path can be
described by the driven diffusion equation
\begin{equation}
{\partial n\over \partial t} = - \mu \, {\rm div} (\bm{F} n) +  D \Delta n - \Gamma n + P_n
\label{eq:diffusion}
\end{equation}
where $n\equiv n(\bm{r},t)$ is the two-dimensional density of the polariton gas, which depends on the 
spatial coordinate $\bm{r}=(x,y)$ in the microcavity plane and time $t$, $\mu$ is the polariton mobility,
$\bm{F}\equiv \bm{F}(\bm{r},t)$ is the external force acting on polaritons, $D$ is the diffusivity, 
$\Gamma$ is the width of the polariton resonance
and $P_n$ is the source (see Ref. \cite{Carusotto:13} for extensive review). In what follows, we
consider the dynamics of the polariton gas in an in-plane trapping potential $U_{trap}(\bm{r}) = {1 \over 2} k r^2$, where 
$k$ is the effective strength of the trap and $r = |\bm{r}|$. The main mechanisms of the diffusive collective motion of  
polaritons include their mutual interactions, scattering on phonons and scettring on excitonic and photonic defects 
in the microcavity \cite{Bley:98}.  (We consider an undoped cavity.)
In our simulations, we take into account the density-dependent polariton interactions in the mean-field approximation
where the effective repulsive interaction potential is $U_{pp} (\bm{r},t)= g n(\bm{r},t)$, which describes exactly the same 
interactions that lead to the blue shift of the polariton excitation spectrum in a superfluid state  \cite{Carusotto:13}.
Here, $g=6 E_B a_B^2 |X_{k}|^4$ is the effective polariton-polariton interaction strength \cite{Carusotto:13},
with $E_B$ and $a_B$ to be the respective binding energy and Bohr radius of the excitons, while $X_{k}$ is the 
Hopfield coefficient that describes the excitonic fraction in the polariton wave function.

In our approach, the driven-diffusive dynamics of the polariton gas is captured via a quasiclassic 
stochastic differential equation for the  center-of-mass coordinate $\bm{r}(t)$ of the polariton wave packets,
\begin{equation}
 d\bm{r}(t) = \mu \bm{F}(\bm{r}(t), t) dt + \sqrt{2D}d\bm{W}(t). \label{eq:stochastic}
\end{equation}
The first term in Eq.\ (\ref{eq:stochastic}) describes the effect of the  force $\bm{F}(\bm{r},t)$ on the 
polariton diffusion and the second term is the Brownian contribution with $d\bm{W}(t)$ to be 
the differential of a Wiener process with  unit variance.  The ensemble-averaged
distribution of ``tracer'' particles  (or quasiparticles)  that obey Eq.\ (\ref{eq:stochastic}) reproduces the  driven-diffusive dynamics (\ref{eq:diffusion}) \cite{Szymczak03}.
The advantage of the model  (\ref{eq:stochastic})  is that, in contrast to  Eq.\ (\ref{eq:diffusion}),  
the stochastic equation describes the fluctuations of the particle density.
The model  (\ref{eq:stochastic}) is off lattice and is computationally effective for systems with a
large number of particles. 
The stochastic model  (\ref{eq:stochastic})
was applied for the description of diffusive properties of quantum gases  (see Ref. \cite{Carusotto:13}) 
as well as for nanoparticle diffusion in fluids \cite{Szymczak03}.

We numerically integrated Eq.\ (\ref{eq:stochastic})  by using  the  Euler method. 
The force acting on polaritons was set equal to $\bm{F} = - \nabla U$ where $U= U_{trap} + U_{pp}$.
The finite-size polariton source at the pumping laser light spot was modeled by adding  particles at each numerical 
time step $\Delta t$ 
with the Gaussian spatial probability distribution centered at the spot center.
 The finite lifetime of polaritons was taken into account by randomly eliminating 
particles from the ensemble at each   time step with the probability $w = \Gamma \Delta t$.

\section{Model parameters}
In our studies, we use the parameters relevant for the A-exciton polaritons 
in a cavity with an embedded
molybdenum disulfide monolayer  \cite{Liu:15}. In our simulations, we  
focus on the polariton dynamics at the room temperature $T=300$ K.

 The effective polariton mass was calculated as
\begin{equation}
m=2/(m_{ex}^{-1} + m_{ph}^{-1}) .
\end{equation}
The exciton mass in MoS$_2$ was taken equal to $m_{ex}=m_e + m_h = 0.78 m_0$ 
where $m_e=0.35 m_0$ and $m_h = 0.43 m_0$ are the electron and hole mass respectively 
(with $m_0$ to be the free electron mass) \cite{Ramasubramaniam:12,Cheiwchanchamnangij:12}.
The effective photon mass is
$m_{ph}={\pi \sqrt{\epsilon} \hbar / c L_C}$,
where the cavity length $L_C$ 
was found from the condition of the resonance of the photonic and excitonic states, 
$E_{ph} = {\pi \hbar c / \sqrt{\epsilon} L_C}$, where $E_{ph}$
is the energy of the cavity photon resonance. The latter was taken equal to $E_{ph}=1.87$ eV \cite{Liu:15}.
The dielectric constant of MoS$_2$ is $\epsilon = 4.3$ \cite{Ramasubramaniam:12}.

\begin{table}[tb]
\centering
\caption{\bf Simulations parameters for a cavity with embedded MoS$_2$ layer}
\begin{tabular}{|lll|}
\hline 
Exciton mass & $m_{ex}$ & $0.78 m_0$ \\
Photon mass & $m_{ph}$ & $1.57 \times 10^{-5} m_0$ \\
Polariton mass & $m$ & $3.14 \times 10^{-5} m_0$ \\
Polariton momentum & $\tau_{pol} $ & $9.60 \times 10^{-13}$ s \vspace{-0.1cm}\\
\hspace{0.15cm} relaxation time & & \\
Polariton diffusion  & $D$ &  $139$ m$^2$/s \vspace{-0.1cm}\\
\hspace{0.15cm} coefficient & & \\
Polariton interaction & $g$ & $1.44 \times 10^{-3}$   \vspace{-0.1cm}\\
\hspace{0.15cm}  strength & &\hspace{0.15cm}  meV$\mu$m$^2$  \\
Polariton mobility & $\mu$ &  $3.36 \times 10^{22}$  \vspace{-0.1cm}\\
 & & \hspace{0.15cm} s/kg \\
Dielectric constant & $\epsilon$ & 4.3 \\
Polariton life time & $\tau_{life}^{(pol)}$ & 84 ps \\
Trapping potential  & $k$ & $15-300$ \vspace{-0.1cm}\\
\hspace{0.15cm} strength & &\hspace{0.15cm}  eV/cm$^2$ \\
Efficiency of polariton & $q $& $10^{-3}$ \vspace{-0.1cm} \\
\hspace{0.15cm} generation & & \\
Numerical unit of  & $\Delta x$ & 1 $\mu$m \vspace{-0.1cm}\\
\hspace{0.15cm}length & & \\
Numerical time step & $\Delta t =$ & 272 fs \vspace{-0.1cm}\\
& \hspace{0.15cm}${m  \Delta x^2}\hbar ^{-1}$ & \\
\hline
\end{tabular}
\label{tab:params}
\end{table}

The exciton diffusion coefficient was taken equal to  $D_{ex} = 14$ cm$^2$/s \cite{Ceballos:15}.  
This value is comparable with the exciton diffusion coefficient in semiconductor quantum wells  \cite{Carusotto:13}.
The polariton  diffusivity was calculated as \cite{Bley:98,Berman:10b}
\begin{equation}
D=|X|^{-4} {m_{ex} \over m} D_{ex}.
\end{equation}
In what follows, we consider zero detuning between the photonic and excitonic resonances thus, we set  $X=1/\sqrt{2}$.
However, the detuning in the experiments with polaritons in a cavity with MoS$_2$ monolayer can be varied in a wide range up to
$\sim 50$ meV \cite{Liu:15}. Thus, the diffusive properties of polaritons can be controlled by
adjusting the detuning value.

The polariton mobility was calculated  as follows \cite{Bley:98,Berman:10b},
\begin{equation}
\mu={\tau_{pol} \over m}.
\end{equation}
The respective momentum relaxation times for excitons  and polaritons  was estimated as
$\tau_{ex}={D_{ex} m_{ex} / k_B T}$ and $\tau_{pol} = |X|^{-4} \tau_{ex}$.

The polariton resonance width was estimated as
\begin{equation}
\Gamma= |X|^2 \Gamma_{ex} + |C|^2  \Gamma_{ph} \label{eq:gamma}
\end{equation}
where $|C|^2 = 1-|X|^2$ is the photon Hopfield coefficient. 
According to Ref.\ \cite{Palummo:15}, the radiative exciton lifetime $\tau_{life}^{(ex)} \equiv \Gamma_{ex}^{-1}$  in MoS$_2$ monolayer
increases from $\tau_{life}^{(ex)} = 0.23$ ps at $T\rightarrow 0$  to $\tau_{life}^{(ex)} \approx 4$ ps at  $T=10$ K and then, to 
270 ps at the room temperature. Even larger radiative exciton life time of 850 ps was reported for MoS$_2$ monolayer crystals at the 
room temperature \cite{Shi:13}.
On the other hand, large broadening of the  excitonic resonance $\hbar \Gamma_{ex} = 30$ meV 
(with the respective time scale of 22 fs) was reported in Ref.\ \cite{Liu:15}.  The respective
short-time processes are probably related to non-radiative  trapping of excitons at impurities or  hot 
carrier thermalization, in agreement with Ref.\ \cite{Palummo:15}.
In our simulation, we set $\tau_{life}^{(ex)} = 270$ ps.
The photon resonance width was taken equal to $\Gamma_{ph} = (50\, {\rm ps})^{-1}$ \cite{Nelsen:13}.
 The polariton life time was estimated  from Eq.\ (\ref{eq:gamma}) as $\Gamma^{-1} = \tau_{life}^{(pol)} =  84$~ps,
 which is in agreement with the observations for semiconductor-based microcavities \cite{Nelsen:13}.

In the simulations, we varied the trapping potential strength from $k=15$ eV/cm$^2$ to 300  eV/cm$^2$ that lies in the experimentally relevant parameter range \cite{Negoita:99,Berman:08}. 
 The  trapping potential can be formed  by applying inhomogeneous 
stress \cite{Negoita:99},  static electric or magnetic field, or laser radiation 
(see Ref.\ \cite{High:12} and references therein).  
 The polariton  interaction constant was calculated as $g=6  E_B a_B^2 |X|^4 = 1.44 \times 10^{-3}$ meV $\mu$m$^2$ 
 given the exciton binding energy $E_B = 0.96$ eV and the exciton Bohr radius $a_B \approx 1$ nm 
 \cite{Liu:15,Shi:13}. It is worth noting that, despite the exciton binding energy $E_B$ in transition metal dichalcogenides
 is significantly larger than 
 that in quantum wells, the polariton interaction strength is of the same order in the both cases.
 
 The polariton density at a point $\bm{r}$ at the moment $t$ 
was calculated from the positions of polariton
quasiparticles $\bm{r}_i (t)$ (where the index $i$ labels the quasiparticles) as follows
\begin{equation}
n(\bm{r},t) = \sum_i f(|\bm{r}_i (t) - \bm{r}|)
\end{equation}
with the weight function $f(r) =  (2\pi r_{av}^2)^{-1} \exp(-r^2 / 2 r_{av}^2)$ and the
averaging length $r_{av} = 10$ $\mu$m. 

Qualitative analysis of experimental data in 
Refs.~\cite{Christopoulos:07,Negoita:99,Feldmann:87,Butov:02a,Malpuech:02a} for quantum well structures
shows that the efficiency of the exciton
generation by a continuous wave (cw) light source is $q\sim 0.1$\%; in other words, only 1 of $10^3$ incident photons creates  
a bound exciton, which is, in its turn, forms a polariton. 
A single-layer transition metal dichalcogenide monolayer can capture up to $10$\% of incident photons \cite{Mak:10,Splendiani:10}. 
However, the efficiency of the  polariton generation for cw pumping in a cavity with MoS$_2$ is not known; we suggest that
the value is close to that obtained for the quantum well structures.
Thus, in our model, 
the polariton injection rate, $P_n$, was calculated as 
\begin{equation}
P_n={q  P  \over E_{ph}}, \label{eq:pn}
\end{equation} 
where $P$ is the pumping laser power.
In the simulations, we consider a Gaussian laser spot profile with full width at half maximum (FWHM) of 30 
$\mu$m. 

The simulations parameters for $T=300$ K are summarized in Table \ref{tab:params}.

\section{Results and Discussion}

\begin{figure}[h]
 \includegraphics[width=8cm]{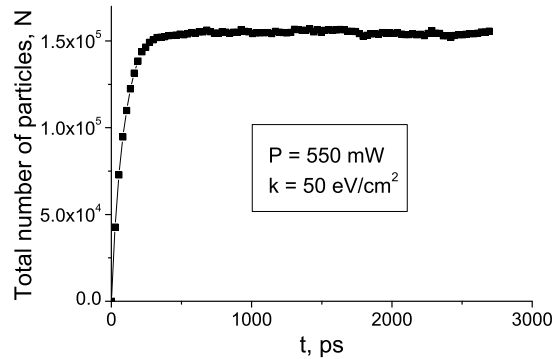}\vspace{-0.5cm}
 \caption{\label{fig:np}
Relaxation to a steady state of a trapped polariton cloud after the laser pump is turned on at the moment $t=0$.
The data points show the total number of polaritons  as a function of time $t$. 
The excitation power of the laser pump is $P=550$ mW, the trapping potential 
strength is  $k=50$ eV/cm$^2$. 
The fluctuations of the total number of polaritons 
in the steady state $t \geq 400$ ps are seen in the figure.}
\end{figure}

First,  we explore the case where the excitation laser spot is positioned at the center of the trap. 
The relaxation of the total number of polaritons in the system to its steady-state value
is shown in Fig.\ \ref{fig:np}. It is seen that the system reaches a steady state  $\sim 400$ ps
after the excitation is switched on.
   
\begin{figure}[tb] 
 \hspace{1 cm } \includegraphics[width=6.5cm]{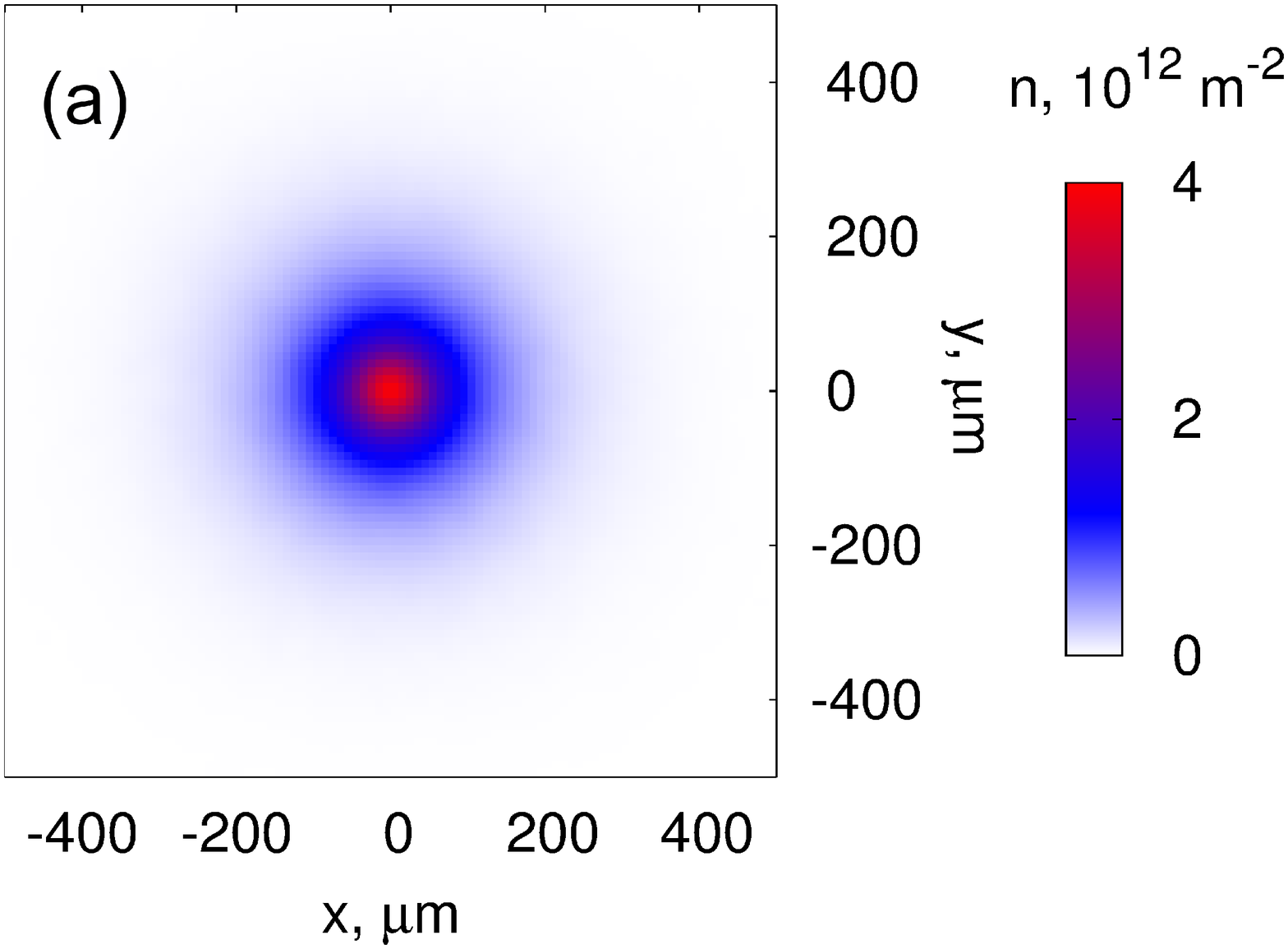}\\
 \includegraphics[width=7cm]{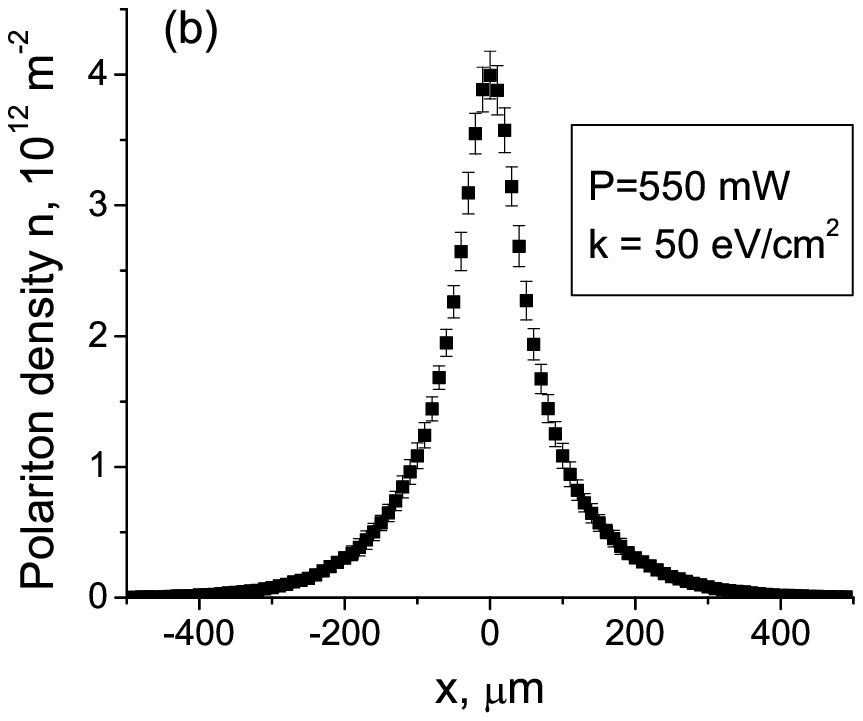}\vspace{-0.5cm}
 \caption{\label{fig:profile} (Color online)
 Spatial distribution of trapped polaritons in a cavity with MoS$_2$ monolayer at $T=300$ K. 
 The power of the laser pump and the trapping potential strength are marked in the figure. 
 (a) The two-dimensional polariton density profile
 $n(\bm{r})$ averaged over the time domain $t=400-2700$ ps, in which  the system has reached the steady state.
 The color bar shows the polariton density in the units of $10^{12}$ m$^{-2}$.
 (b) A one-dimensional cut of the graph (a) made along the line  $y=0$.
 Points show the results of the simulations, 
 vertical bars show the standard deviation about the mean for the density fluctuations. 
 In the simulations, the full width at half maximum (FWHM) of the laser spot is 30 $\mu$m; the center of the laser spot
is positioned at the center of the parabolic trap  $\bm{r}=(0,0)$. }
\end{figure}

The steady-state spatial distribution of polaritons in the trap for the trapping potential strength $k=50$ eV/cm$^2$ is shown in Fig.\ \ref{fig:profile}.
It is seen that the characteristic polariton cloud size reaches $\sim 200$ $\mu$m that is, about an order larger than the excitation
spot size 30 $\mu$m.
It is also much larger than the Thomas-Fermi size of a cloud at zero temperature 
$a_{TF}=\sqrt{2 n_{max} g / k}\approx 1-10$ $\mu$m for the maximum density at the center  
$n_{max} \approx n_c$ \cite{Berman:08}.
 The main reason for cloud spreading is thermal diffusion of polaritons towards the edges of 
the cloud from the excitation area. Thus, at room temperatures, 
 thermal  spreading   is more pronounced than the  effect of the interparticle interactions. 

\begin{figure}[h] 
\centerline{ \includegraphics[width=8cm]{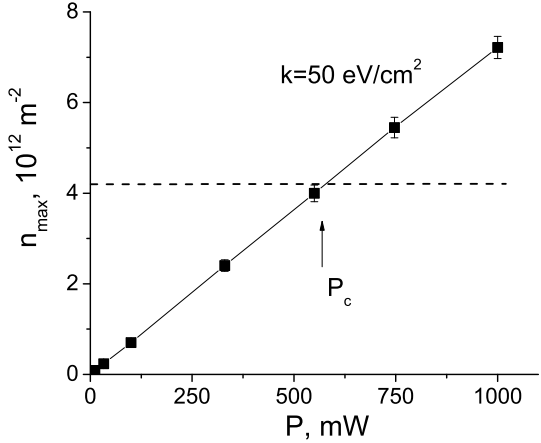}}\vspace{-0.5cm}
 \caption{\label{fig:nmax}
Increase in the maximum density of polaritons, $n_{max}$, at the center of the cloud $\bm{r}=(0,0)$
with the rise of the power of the laser pump $P$.
Points show the results of the simulations for $k=50$ eV/cm$^2$, the line segments connecting the points 
are shown to guide the eye.
The dashed horizontal line corresponds to the critical density $n_c=4.2 \times 10^{12}$ m$^{-2}$,
at which a gas of interacting polaritons in a cavity with MoS$_2$ undergoes the Kosterlitz-Thouless superfluid transition
at the room temperature. 
A polaritons gas is superfluid at the central part of the cloud 
above the line ($P>P_c \approx 600$ mW) and  is in the normal state below the line ($P<P_c$).}
\end{figure}

Figure \ref{fig:nmax} shows the maximum density of polaritons, $n_{max}$, reached 
at the center of the cloud as a function of the excitation
power $P$. The data are shown for the trapping potential strength  $k=50$ eV/cm$^2$ and temperature $T=300$ K.
It is seen that the maximum density $n_{max}$ nearly linearly grows with the increase of the power $P$.
For the excitation power $P=P_c \approx 600$ mW, the maximum density of polaritons reaches 
the critical density $n_c\approx 4.2\times 10^{12}$ m$^{-2}$. At the pumping power $P>P_c$ the central part 
of a polariton system at room temperatures should demonstrate superfluid properties previously known for 
a low-temperature polariton superfluid in a semiconductor microcavity (see review \cite{Carusotto:13}).
The wings of the spatial distribution always have a density lower than $n_c$ thus,
the superfluid ``lake'' at the cloud center is surrounded by a low-density 
gas of polaritons in a normal state.

We emphasize that, after a part of the polariton system transits to a superfluid state,
quantum correlations in the polariton dynamics should be taken into account.
However, if the superfluid density is small compared to the total density of the system, 
$n_{max} - n_c < n_c$,
the superfluid effects on the over-all large-scale density distribution in a steady state  can be disregarded 
to the first approximation. In particular, the presence of a small superfluid fraction
will not  modify the value of the  maximum density in the cloud and thus,
the criterion of the superfluidity formation will not be changed.
(We note that a few-percent density jump can occur during the superfluid transition, in a similarity with the 
normal-to-superfluid transition in He-II \cite{Khalatnikov:65}; in our case however, this jump 
is less than or comparable  with a typical value of the thermal  fluctuations of the density $\sim 4$\% estimated from 
Fig.\ \ref{fig:profile} and therefore, it can be disregarded.) 
We also disregard the dissipative dynamics at a very close vicinity of the superfluid transition
where the superfluid component can appear as a fluctuation mode \cite{Khalatnikov:72}.
Our goal is to pinpoint a domain in the parameter space where the room-temperature superfluidity 
of polaritons can be experimentally achieved.

\begin{figure}[tb] 
 \centerline{ \includegraphics[width=8cm]{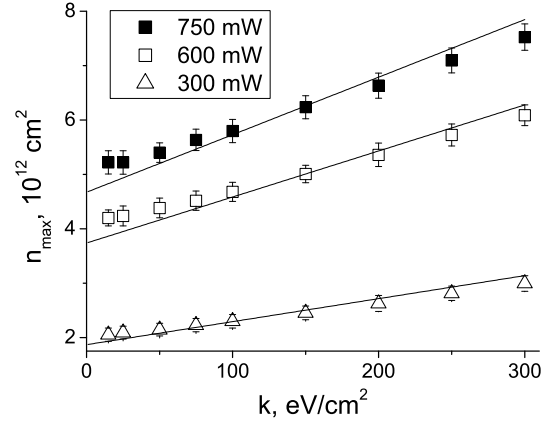}}\vspace{-0.5cm}
 \caption{\label{fig:nk}
Dependence of the maximum density of the polariton cloud on the trapping potential strength $k$.
Points show the results of the simulations, 
lines show the theoretical dependence Eq.\ (\ref{eq:mnax}) for the respective parameters. 
The  values for the laser pump power are listed in the figure.}
\end{figure}

To better understand the optimal conditions for observations of room-temperature polariton superfluidity,
we studied the dependence of the maximum density  $n_{max}$ on the trapping potential strength. 
The results of the numerical simulations are summarized in Fig.\ \ref{fig:nk}.
It is seen that the density  grows with the rise of the trapping potential strength.
From the obtained results it follows that at experimentally realistic pumping power 
 $P \sim 600$  mW, the cloud can transit to a superfluid state at $T=300$ K for the trapping strength $k \geq 50$ eV/cm$^2$.
 For a larger power $P = 750$ mW, the polariton density required for the room-temperature superfluidity 
 is reached at even  smaller trapping strengths $k\geq 15$ eV/cm$^2$. 
 On the other hand, for $P = 300$ mW, the conditions for the room temperature polariton superfluidity 
 are not satisfied even for large trapping strengths $k \sim 300$ eV/cm$^2$.  
 
\begin{figure}[h] 
 \hspace{1.5 cm } \includegraphics[width=6cm]{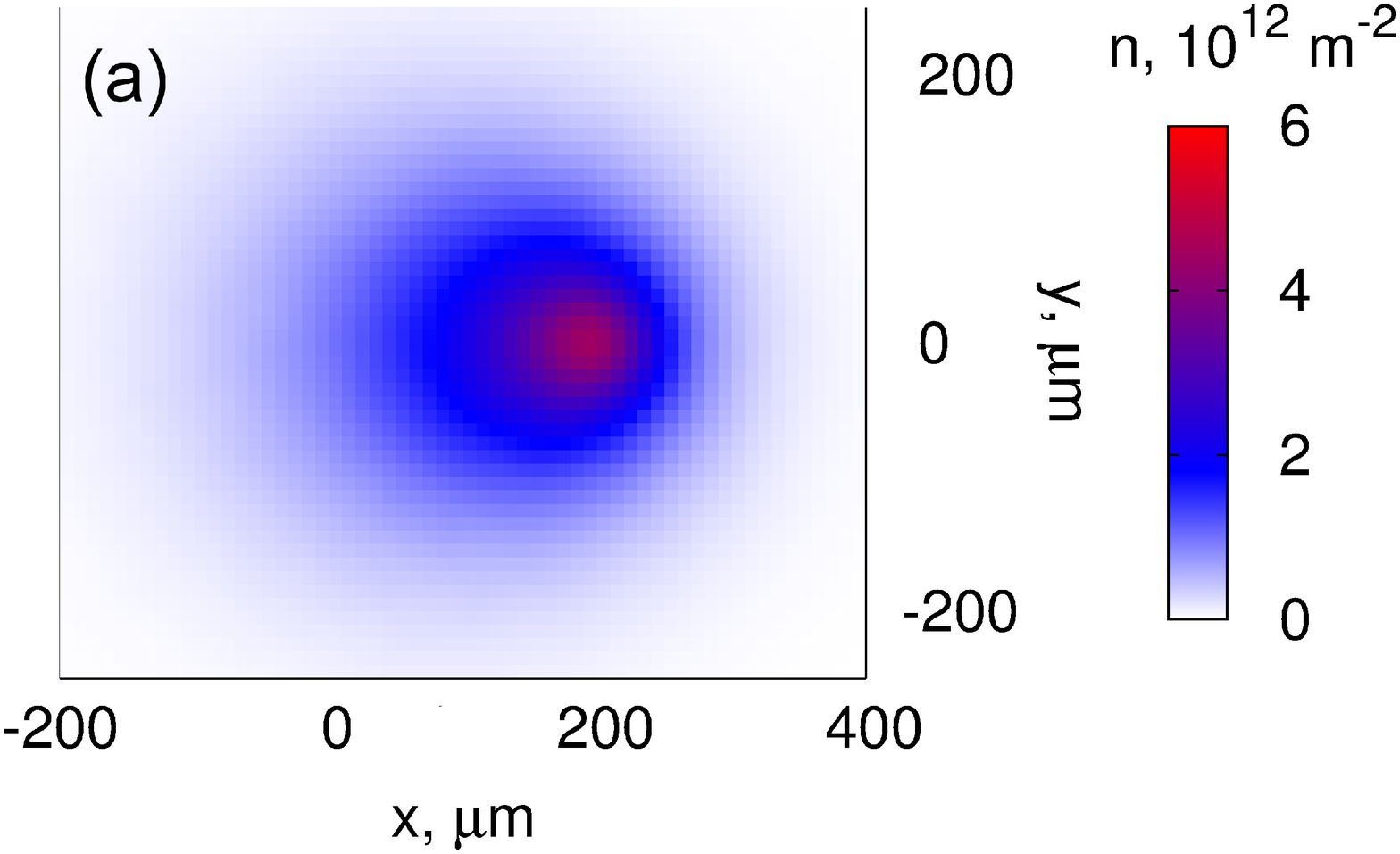}\\
 \includegraphics[width=7cm]{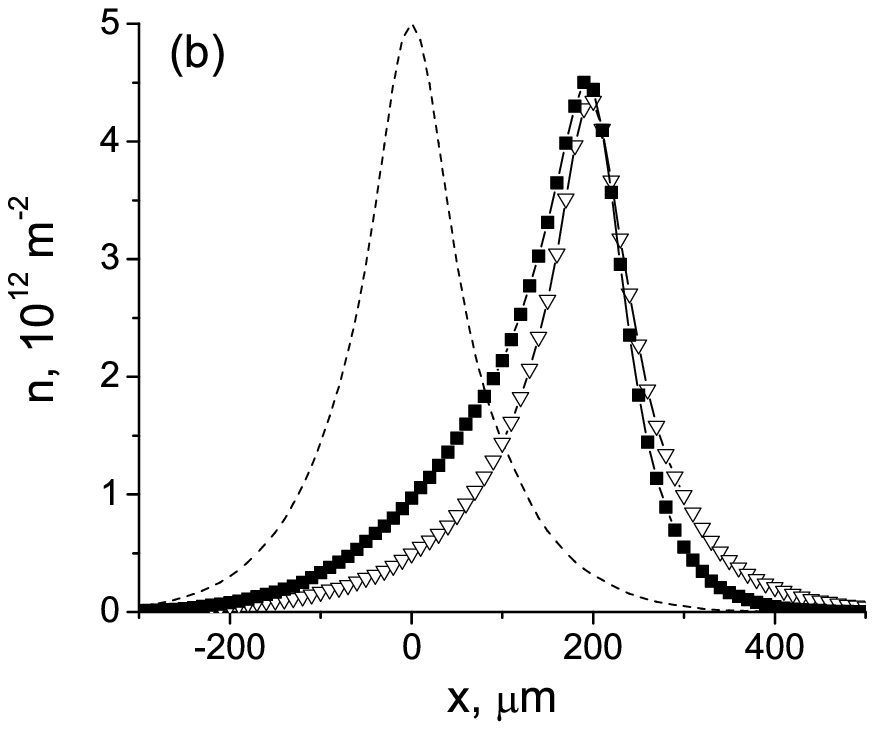}\vspace{-0.5cm}
 \caption{\label{fig:offset} (Color online)
Polariton diffusion towards the center of the trap  in the case where the excitation 
spot is shifted from the trap center. The pumping power is $P=600$ mW.
The offset for the excitation spot center is $x_0=200$ $\mu$m in the positive direction of the 
$x$ axis. (a) Two-dimensional polariton density distribution for $k=150$ eV/cm$^2$.
(b) The density distribution calculated along the line $y=0$ for
$k=50$ (unfilled triangles) and 150 eV/cm$^2$ (filled squares). For the reference, the density distribution for
$k=50$ eV/cm$^2$ with zero excitation spot offset, $x_0=0$, is shown by the dashed curve. 
Fluctuations of the polariton density ($\sim 4$\%) are not shown.}
\end{figure}

To further investigate the trapped polariton gas dynamics, we also studied 
the effect of the spatial offset in the excitation spot position compared to the
center of the parabolic trap.  For this purpose, we numerically studied the polariton spatial distribution
when the center of the laser spot is shifted along the $x$ axis on the distance 
$x_0=200$ $\mu$m that is, larger than the excitation spot size of 30 $\mu$m.
The obtained numerical results are shown in Fig.\ \ref{fig:offset}.
Fig.\ \ref{fig:offset}a shows the two-dimensional polariton density distribution for $k=150$ eV/cm$^2$, and 
Fig.\ \ref{fig:offset}b shows its cross-section made along the $x$ axis at $y=0$ for $k=150$ eV/cm$^2$
(filled squares) and for $k=50$ eV/cm$^2$ (unfilled triangles).  
It is seen that the polaritons diffuse towards the center of the trap $\bm{r}=0$.
It is also seen that the polariton density at the center of the trap 
increases with the rise of the trapping potential strength $k$. However, for moderate 
strengths $k \leq 150$ eV/cm$^2$, the polariton density at $\bm{r}=0$
remains small compared to the maximum density at the excitation spot area.
Specifically, the density at $\bm{r}=0$ is $\approx 19$\% of the maximum density in the cloud for $k=150$ eV/cm$^2$
and it is only 10\% for $k=50$ eV/cm$^2$.

The geometry with a shifted excitation spot might provide an additional opportunity for the experimental detection of
room-temperature polariton superfluidity. One can expect a larger occupation of the lowest energetic state 
near the trap center  $\bm{r}=0$ for a superfluid gas compared to the case of a normal gas (these simulations).
In other words, the superfluid polariton gas would ``easier'' flow towards the bottom of the trap 
that, in its turn, should lead to higher densities $n(0)$.
Thus, observation of  ``anomalously'' high polariton population $n(0)$ 
with the increase of the trapping strength $k$ or the laser power $P$
in experiments with a shifted laser spot can be a simple experimental test of the
normal-to-superfluid transition in the polariton gas.

Finally, we provide an estimate of the  cloud densities that facilitates 
the search of  parameters suitable for the polariton superfluidity observation.
As it follows from the above consideration, a key parameter that determines the feasibility of polariton superfluidity
at room temperatures is the maximum polariton density at the center of the cloud $n_{max}$.
To estimate the maximum density $n_{max}$  the following  model can be utilized.
In the case of small trapping potential strengths, the limiting factor for the  cloud size is
a finite polariton life time $\tau_{life}^{(pol)}$.  In this limit, the cloud size can be estimated as 
$a=(D \tau_{life}^{(pol)})^{1/2}$. In the opposite case of strong trapping potentials, the 
cloud size is limited by thermal fluctuations, $k_B T \sim {1 \over 2} k a^2$, and thus, it 
can be estimated as $a = (2 k_B T / k)^{1/2}$. In the intermediate case, the
cloud size can be obtained from the interpolation equation
\begin{equation}
a = {(D\tau_{life}^{(pol)})^{1/2} \over \sqrt{ 1 + \gamma}} \label{eq:a}
\end{equation} 
where $\gamma = k D\tau_{life}^{(pol)} / 2 k_B T$ is the dimensionless trapping potential strength.
We note that for the parameters corresponding to an embedded MoS$_2$ monolayer at $T=300$ K (Table \ref{tab:params}), one obtains
 $\gamma = 1$ for $k \approx 440$ eV/cm$^2$.
By approximating the steady-state cloud shape as an axially-symmetric  profile $n(r) = n_{max} /(1+(r/ a)^6)$
and the total number of polaritons in the system as $N \approx P_n \tau_{life}^{(pol)}$, one has the following 
expression for the maximum polariton density, 
\begin{equation}
 n_{max} = \frac{3 \sqrt{3}}{2 \pi ^2} {P_n \tau_{life}^{(pol)} \over  a^2}
 \approx 0.26 {P_n \tau_{life}^{(pol)} \over  a^2}, \label{eq:mnax}
\end{equation}
in which the parameter $a$ is given in Eq.\ (\ref{eq:a}) and the source $P_n$ is estimated in Eq.\ (\ref{eq:pn}).
The model Eq.\ (\ref{eq:mnax}) is in good agreement with the results of the simulations, as is
seen in Fig.~\ref{fig:nk}.

\section{conclusions}
In this paper, we studied the polariton gas dynamics in a parabolic trapping potential by using the quasiclassical
stochastic equation for the polarion wave packets diffusion.
 We showed that in a cavity with an embedded molybdenum disulfide monolayer, 
 room-temperature superfluidity in a trapped polariton  
cloud can be observed for experimentally 
realistic trapping strengths $k \sim 50-300$ eV/cm$^2$ and laser excitation power $P\geq 600$ mW.
The experiments with a shifted laser spot would provide a simple qualitative  test
for superfluidity. 
The theoretical estimate  shows that
a trapped polariton gas  is superfluid at the central part of the polariton cloud if
the laser pumping power exceeds  the critical value
\begin{equation}
P_c = {3.8 D n_c \over 1+\gamma} {E_{ph} \over q}.
\end{equation}

%We also proposed a simple model that facilitate the parameter estimates for the 
%experimental search of room-temperature polariton superfluidity.

In our studies, we are inspired by the recent observation of room temperature polaritons \cite{Liu:15}
as well as by the advances of low-temperature polariton physics \cite{Carusotto:13,Gibbs:11}.
Superfluidity of excitons in transition metal dichalcogenide-based systems
has recently been predicted for temperatures beyond the helium temperature range \cite{Berman:16}.
In addition to the fundamental importance of high-temperature quantum effects,
polariton superfluidity might be a key to  new optoelectronic  technologies including  
 those  mentioned in Introduction.
Room-temperature polariton superfluidity would open a new route towards the design of 
photonic integrated circuits suitable for the use in optical information processing systems with low energy consumption.
Our hope is that the experimental search for room-temperature quantum  fundamental effects 
in transition metal dichalcogenide-based heterostructures would facilitate the progress in this direction.

\section{Funding Information}
National Science Foundation (NSF) (Supplement to 1345219); 
Professional Staff Congress -- City University of New York (PSC-CUNY)  (68090-00 46).

The authors thank the Center for Theoretical Physics of the New York City College of 
Technology for providing computational resources.

% Bibliography
%on laptop:
%\bibliography{/Users/kolmakov/Dropbox/Docs/Papers/lowtempgk,/Users/kolmakov/Dropbox/Docs/Papers/tmd}
%desktop:
%\bibliography{/home/kolmakov/Dropbox/Docs/Papers/lowtempgk,/home/kolmakov/Dropbox/Docs/Papers/tmd}
%merlin.mbs apsrev4-1.bst 2010-07-25 4.21a (PWD, AO, DPC) hacked
%Control: key (0)
%Control: author (0) dotless jnrlst
%Control: editor formatted (1) identically to author
%Control: production of article title (0) allowed
%Control: page (1) range
%Control: year (0) verbatim
%Control: production of eprint (0) enabled
%

\end{document}